\documentclass[aps,prd,floatfix,notitlepage,superscriptaddress,preprintnumbers,nofootinbib,11pt]{revtex4-2}

\usepackage{graphicx} % Required for inserting images

\usepackage[T1]{fontenc} % if needed
\usepackage{comment}
\usepackage{appendix}
\usepackage{hyperref}

\usepackage[dvipsnames]{xcolor}
\usepackage{feynmp-auto}
\usepackage{mathtools,slashed}
\usepackage{tikz}
\usepackage{amsmath}
\usepackage{amssymb}
\usepackage{cancel}
\usepackage{soul} % stikethrough text
\usepackage{multirow}

\newcommand\Tr{\text{Tr}}

\begin{document}
\title{Revisiting the two-zero texture Majorana neutrino mass matrix}

\author{Wararat Treesukrat}
\email{w.treesukrat@gmail.com}
\affiliation{Theoretical High-Energy Physics and Astrophysics Research Unit, Department of Physics, Srinakharinwirot University, 114 Sukhumvit 23 Rd., Wattana, Bangkok
10110, Thailand}

\author{Nopmanee Supanam}
\email{nopmanee@g.swu.ac.th}
\affiliation{Theoretical High-Energy Physics and Astrophysics Research Unit, Department of Physics, Srinakharinwirot University, 114 Sukhumvit 23 Rd., Wattana, Bangkok
10110, Thailand}

\author{Patipan Uttayarat}
\email{patipan@g.swu.ac.th}
\affiliation{Theoretical High-Energy Physics and Astrophysics Research Unit, Department of Physics, Srinakharinwirot University, 114 Sukhumvit 23 Road, Wattana, Bangkok
10110, Thailand}

\begin{abstract}
It has long been pointed out that there are seven different two-zero texture neutrino mass matrices compatible with neutrino oscillation data. We perform an updated analysis with the recently published Nu-Fit 6.0 results. We also subject the seven two-zero textures to constraints on neutrino mass from cosmology, end point of beta decay spectrum, and neutrinoless double beta decay experiments. We find that all seven textures are compatible with the new oscillation parameters. However, we find five textures, whose 1-1 entry is nonvanishing, are in severe tension with the constraints from cosmology and neutrinoless double beta decay. With the next generation experiments, these five textures could be decisively ruled out. For the remaining two textures, one of them could be ruled out, or severely constrained, if the octant of $\theta_{23}$ is determined. 
\end{abstract}

\maketitle
\flushbottom

%%%%%%%%%%%%%%%%%%%%
\section{Introduction}
The discoveries of neutrino oscillations have established the massive nature of neutrinos. For the three active neutrinos, their oscillations can be described by six real parameters: two mass-squared differences $\Delta m^2_{\text{sol}}$ and $\Delta m^2_{\text{atm}}$; three mixing angles $\theta_{12}$, $\theta_{23}$, and $\theta_{13}$; and a $CP$ violating phase $\delta$ commonly referred to as the Dirac phase. Global analysis of neutrino oscillation data has determined the values of $\Delta m^2_{\text{sol}}$,  $\Delta m^2_{\text{atm}}$, $\theta_{12}$, $\theta_{23}$, and $\theta_{13}$ to a high degree of precision~\cite{deSalas:2020pgw,Capozzi:2021fjo,Esteban:2024eli}. The Dirac phase, on the other hand, has not been as precisely determined. Even though we know quite a lot about the parameters governing neutrino oscillations, our knowledge of the nature of neutrino masses is far from complete. We do not know the absolute neutrino mass scale nor their ordering. We also do not know if neutrinos are Dirac or Majorana particles. Although a Dirac neutrino is a concrete possibility, in this work, we will solely focus on the Majorana neutrino. 

There is a plethora of neutrino mass models predicting the Majorana neutrino. Prototypical examples are the seesaw models~\cite{Minkowski:1977sc,Yanagida:1979as,GellMann:1980vs,Mohapatra:1979ia} and the models where neutrino masses are generated radiatively~\cite{Zee:1980ai,Zee:1985id,Babu:1988ki}. Because of the lack of experimental hints for any of the neutrino mass models, we will remain agnostic to the mechanism of neutrino mass generation. We work instead with the low-energy effective Majorana neutrino mass matrix which is a 3$\times$3 complex symmetric matrix. Such a matrix contains 12 real parameters, three of which can be absorbed into the phase of the lepton fields. The remaining nine parameters are physical. Unfortunately, they cannot be fully determined from the six available neutrino oscillation parameters. 

To reduce the number of free parameters in the Majorana mass matrix, Refs.~\cite{Xing:2002ta,Frampton:2002yf} have introduced an ansatz where some entries of the mass matrix $m_\nu$, in the charged lepton mass eigenbasis, are vanishing. It is easy to see that $m_\nu$ which has more than two independent entries being zero is incompatible with the oscillation data. As a result, the case of $m_\nu$ with two independent entries vanishing, often referred to as two-zero texture, has gained a lot of attention in the literature. A two-zero texture $m_\nu$ contains five free parameters, one less than the number of available neutrino oscillation parameters. As a result, the two-zero texture ansatz can be predictive. In principle, there are 15 different two-zero textures. However, as of 2011, only seven of them are compatible with neutrino data~\cite{Frampton:2002yf,Guo:2002ei,Fritzsch:2011qv,Kageyama:2002zw,Frigerio:2002fb,Honda:2003pg,Bhattacharyya:2002aq,Mohanta:2006xd,Farzan:2006vj,Dev:2009he,Dev:2006qe,Lashin:2007dm,Dighe:2009xj,Goswami:2008rt,Lashin:2009yd,Ahuja:2009jj,Goswami:2009bd,Dev:2010pe,Dev:2010if,Grimus:2011sf}. Adopting the notation of Ref.~\cite{Frampton:2002yf}, they are
% {\color{blue} ~\cite{Kageyama:2002zw,Frigerio:2002fb,Honda:2003pg,Bhattacharyya:2002aq,Mohanta:2006xd,Farzan:2006vj,Dev:2009he,Dev:2006qe,Lashin:2007dm,Dighe:2009xj,Goswami:2008rt,Lashin:2009yd,Ahuja:2009jj,Goswami:2009bd,Dev:2010pe,Dev:2010if,Grimus:2011sf}[Add more references. maybe add all in Ref [14] of 1108.4534]}. Adopting the notation of Ref.~\cite{Frampton:2002yf}, they are
\begin{align*}
    &\mathrm{A1:}\: \begin{pmatrix}0 &0 &\ast\\ &\ast &\ast\\&&\ast \end{pmatrix},\qquad
    &\mathrm{A2:}\: \begin{pmatrix}0 &\ast &0\\ &\ast &\ast\\&&\ast \end{pmatrix};\\
    &\mathrm{B1:}\: \begin{pmatrix}\ast &\ast &0\\ &0 &\ast\\&&\ast \end{pmatrix},\qquad
    &\mathrm{B2:}\: \begin{pmatrix}\ast &0 &\ast\\ &\ast &\ast\\&&0 \end{pmatrix},\\
    &\mathrm{B3:}\: \begin{pmatrix}\ast &0 &\ast\\ &0 &\ast\\&&\ast \end{pmatrix},\qquad
    &\mathrm{B4:}\: \begin{pmatrix}\ast &\ast &0\\ &\ast &\ast\\&&0 \end{pmatrix};
    \\
    &\mathrm{C:}\phantom{1}\: \begin{pmatrix}\ast &\ast &\ast\\ &0 &\ast\\&&0 \end{pmatrix},
\end{align*}
where $\ast$ denotes a nonzero entry. Note that the elements below the main diagonal are related to those above the main diagonal by symmetricity of the mass matrix. 

The low-energy two-zero texture Majorana neutrino mass matrix can arise from many neutrino mass models. Reference~\cite{Fritzsch:2011qv} shows that the seesaw model with a discrete $Z_N$ flavor symmetry can generate the two-zero texture. Moreover, the Zee model for neutrino mass with a minimal Yukawa structure also predicts two-zero texture neutrino mass matrices~\cite{Primulando:2022vip}. In fact, such a minimal Zee model can also explain the muon $g-2$ anomaly~\cite{Primulando:2022vip}. 

In this work, we perform an updated analysis of the seven two-zero textures to determine the extent to which they are compatible with the current neutrino data. We systematically scan over the full parameter space of the two-zero texture mass matrix. As a result of the scan, we also determine the possible ranges of the lightest neutrino mass ($m_{\mathrm{light}}$), the effective masses for neutrinoless double beta decay experiment ($m_{ee}$), and the effective mass describing the end point of the beta decay spectrum ($m_{\nu_e}^{\mathrm{eff}}$), for each of the seven two-zero textures. This allows us to exclude a large portion of parameter space of some of the textures.

The paper is organized as follows. In Sec.~\ref{sec:notation} we introduce the seven two-zero textures to set up our notations and conventions. We then work out the phenomenology of the seven textures in Sec.~\ref{sec:pheno}. The results of numerical scan on the parameter space of the seven textures are presented in Sec.~\ref{sec:results}. Finally, we conclude in Sec.~\ref{sec:conclusion}.

%%%%%%%%%%%%%%%%%%%%
\section{Two-zero texture neutrino mass matrix}
\label{sec:notation}
The two-zero texture neutrino mass matrix, in the basis where charged leptons are diagonal, contains four complex independent elements, thus, eight real parameters.  However, not all of them are physical. Under phase redefinitions   
\begin{equation}
    L_j \to e^{i\alpha_j}L_j,\quad e_{R_j}\to e^{i\alpha_j}e_{R_j},
\end{equation}
where $L$ is the left-hand lepton doublet and $e_R$ is the right-hand lepton singlet, the charged lepton mass matrix remains invariant while 
\begin{equation}
    (m_\nu)_{jk} \to e^{i(\alpha_j+\alpha_k)}(m_\nu)_{jk}.
\end{equation}
The above phase transformations allow one to eliminate three phases in $m_\nu$. As a result, $m_\nu$ can be parametrized by five real parameters. 

In this work, we parametrize the seven two-zero textures as follows:
\begin{align}
    m_\nu^\mathrm{(A1)} &= \mu\begin{pmatrix}0 &0 &r_2\\ 0&e^{i\theta} &r_3\\r_2&r_3&r_1 \end{pmatrix},\qquad
    m_\nu^\mathrm{(A2)} = \mu\begin{pmatrix}0 &r_2 &0\\ r_2&r_1 &r_3\\0&r_3&e^{i\theta} \end{pmatrix};\label{eq:textureA}\\
    m_\nu^\mathrm{(B1)} &= \mu\begin{pmatrix}e^{i\theta} &r_2 &0\\ r_2&0 &r_3\\0&r_3&r_1 \end{pmatrix},\qquad
    m_\nu^\mathrm{(B2)} = \mu\begin{pmatrix}e^{i\theta} &0 &r_2\\ 0&r_1 &r_3\\r_2&r_3&0 \end{pmatrix},\nonumber\\
    m_\nu^\mathrm{(B3)} &= \mu\begin{pmatrix}e^{i\theta} &0 &r_2\\ 0&0 &r_3\\r_2&r_3&r_1 \end{pmatrix},\qquad
    m_\nu^\mathrm{(B4)} = \mu\begin{pmatrix}e^{i\theta} &r_2 &0\\ r_2&r_1 &r_3\\0&r_3&0 \end{pmatrix};\label{eq:textureB}\\
    m_\nu^\mathrm{(C)} &= \mu\begin{pmatrix}e^{i\theta} &r_1 &r_2\\ r_1&0 &r_3\\r_2&r_3&0 \end{pmatrix}\label{eq:textureC};
\end{align}
where $\mu$ is the scale of neutrino masses, $r_i>0$ is the dimensionless ratio, and $\theta$ is the phase. The ratio $r_i$ is expected to be of $\mathcal{O}(1)$. If $r_i\gg1$ or $r_i\ll1$, the neutrino mass matrix will have approximately three or more independent zeros, rendering it incompatible with neutrino oscillation data. 

There are three basic invariants that can be constructed from the above neutrino mass matrix. They are
\begin{align}
    \Tr(m_\nu^\dagger m_\nu^{}) &= m_1^2 + m_2^2 + m_3^2,\label{eq:trace}\\
    \frac{1}{2}\left((\Tr[m_\nu^\dagger m_\nu^{}])^2 - \Tr[(m_\nu^\dagger m_\nu^{})^2]\right) &= m_1^2m_2^2 + m_1^2m_3^2 + m_2^2m_3^2,\label{eq:cross}\\
    \det(m_\nu^\dagger m_\nu) &= m_1^2m_2^2m_3^2\label{eq:det},
\end{align}
where $m_i^2$ is the mass squared of neutrino mass eigenstate $\nu_i$. In terms of our parametrization above, these invariants take the same form for textures A1 and A2, B1 and B2, and B3 and B4. As a result, we expect neutrino masses to show similar behavior within each pair of textures.

%%%%%%%%%%%%%%%%%%%%
\section{Neutrino phenomenology}
\label{sec:pheno}
In this section, we study the neutrino phenomenology of the seven two-zero texture neutrino mass matrices. The five parameters describing the two-zero textures [see Eqs.~\eqref{eq:textureA} -~\eqref{eq:textureC}] can be determined by fitting to the six neutrino oscillation parameters: three mixing angles $s_{12}^2$, $s_{23}^2$, and $s_{13}^2$; two mass-squared splittings $\Delta m^2_{\mathrm{sol}}$ and $\Delta m^2_{\mathrm{atm}}$; and the Dirac phase $\delta$. The values of neutrino oscillation parameters, determined from the global analysis of neutrino data~\cite{Esteban:2024eli}, are collected in Table~\ref{table:data_nufit}.

The above neutrino mass matrix can be diagonalized by
\begin{equation}
    U^Tm_\nu U = \text{diag.}\left(m_1,m_2,m_3\right),
    \label{eq:diagonalization}
\end{equation}
where $m_i$ is real and non-negative. Equation~\eqref{eq:diagonalization} defines the Pontecorvo-Maki-Nakagawa-Sakata (PMNS) matrix $U$. In the literature, the PMNS matrix is commonly parametrized as
\begin{equation}
    U = \begin{pmatrix}1&0&0\\0&c_{23}&s_{23}\\0&-s_{23}&c_{23} \end{pmatrix}
    \begin{pmatrix}c_{13}&0&s_{13}e^{-i\delta}\\0&1&0\\-s_{13}e^{i\delta}&0&c_{13}\end{pmatrix}
    \begin{pmatrix}c_{12}&s_{12}&0\\-s_{12}&c_{12}&0\\0&0&1 \end{pmatrix}
    \begin{pmatrix}e^{i\eta_1/2}&0&0\\0&e^{i\eta_2/2}&0\\0&0&1 \end{pmatrix},
\end{equation}
where $s_{ij}$ ($c_{ij}$) is the sine (cosine) of the lepton mixing angle $\theta_{ij}$, $\delta$ is the Dirac phase, and $\eta_1$ and $\eta_2$ are Majorana phases. Note the factor of 1/2 in the definition of Majorana phases is introduced so that both Dirac and Majorana phases can be taken in the range [0,$2\pi$]~\cite{Jenkins:2007ip}.

After diagonalizing the neutrino mass matrix, the mixing angles can be determined from
\begin{equation}
    s_{12}^2 = \frac{|U_{e2}|^2}{1-|U_{e3}|^2},\quad
    s_{13}^2 = |U_{e3}|^2,\quad
    s_{23}^2 = \frac{|U_{\mu3}|^2}{1-|U_{e3}|^2}.
\end{equation}
The mass-squared splittings are defined as $\Delta m^2_{\text{sol}} = m_2^2-m_1^2$, and $\Delta m^2_{\text{atm}} = m_3^2-m_1^2$ and $m_2^2-m_3^2$ for normal ordering (NO) and inverted ordering (IO) neutrino masses, respectively. The sine of the Dirac phase can be deduced from the Jarsklog invariant
\begin{equation}
    J_{CP} = \text{Im}\left[U_{e2}^{}U_{\mu2}^*U_{\mu3}^{}U_{e3}^*\right].
\end{equation}
Finally, the two Majorana phases $\eta_1$ and $\eta_2$ can be deduced from the quadratic invariants $U_{e1}^{}U_{e2}^*$ and $U_{e1}^{}U_{e3}^*$~\cite{Nieves:2001fc,Jenkins:2007ip}. The former is proportional to $e^{i(\eta_1-\eta_2)/2}$, while the latter is proportional to $e^{-i(\delta-\eta_1/2)}$. However, both Majorana phases have never been determined experimentally.

\begin{table}[ht]
    \centering
    \begin{tabular}{|l|c|c|c|c|c|l|}
    \hline\hline
\multirow{2}{*}{Parameters}  & \multicolumn{2}{c|}{Normal ordering}  & \multicolumn{2}{c|}{Inverted ordering} \\\cline{2-5}
    &Best fit value & $3\sigma$ range &  Best fit value & $3\sigma$ range \\ \hline
$\sin^2\theta_{12}$ & $0.308_{-0.011}^{+0.012}$  & $0.275 \to 0.345$ & $0.308_{-0.011}^{+0.012}$ & $0.275 \to 0.345$ \\
$\sin^2\theta_{23}$ & $0.470_{-0.013}^{+0.017}$ & $0.435 \to 0.585$ & $0.550_{-0.015}^{+0.012}$ & $0.440 \to 0.584$\\
% $\left[\sin^2\theta_{23} \text{(shallow)}\right]$ & $0.550_{-0.045}^{+0.014}$ & $0.416 \to 0.592$ & $ 0.470_{-0.012}^{+0.040}$ & $0.435 \to 0.591$ \\
$\sin^2\theta_{13}$ & $0.02215_{-0.00058}^{+0.00056}$ & $0.02030 \to 0.02388$ & $0.02231_{-0.00056}^{+0.00056}$ & $0.02060 \to 0.02409$ \\
\hline 
$\Delta m_{\mathrm{sol}}^2/10^{-5}$ eV$^2$ & $7.49_{-0.19}^{+0.19}$ & $6.92\to 8.05$ & $7.49_{-0.19}^{+0.19}$ & $6.92\to 8.05$\\
        $\Delta m_{\mathrm{atm}}^2/10^{-3}$ eV$^2$ & $2.513_{-0.019}^{+0.021}$ & $2.451\to 2.578$ & $2.484_{-0.020}^{+0.020}$ & $ 2.421\to2.547$ \\
        \hline
        $\delta_{\mathrm{CP}}/^{\circ}$ & $212_{-41}^{+26}$ & $ 124\to 364$  & $ 274_{-25}^{+22}$  & $201 \to 335$ \\
        % \hline
        % $R$ &  &  &  &\\
        %\hline
        %$J_{\mathrm{CP}}$  & $-0.0178$  &  & $-0.0334$ & \\
\hline
\hline
\end{tabular}
    \caption{Best fit values and the $3\sigma$ ranges for neutrino oscillation parameters obtained from NuFIT 6.0~\cite{Esteban:2024eli}. Here, $\Delta m_{\mathrm{sol}}^2=m_2^2-m_1^2$ and $\Delta m_{\mathrm{atm}}^2=m_3^2-m_1^2$ and $m_2^2-m_3^2$ for normal ordering and inverted ordering, respectively. 
    %For $\sin^2\theta_{23}$ we allow for the possibility that $\theta_{23}$ lies in the shallow minimum.
    %{\color{blue}[PU: to simplify, we dont use shallow minimum for $s_{23}^2$. We will use instead the 3$\sigma$ range which should more or less cover the shallow minimum. Also, still missing entry for Dirac phase $\delta$]}
    }
    \label{table:data_nufit}
\end{table}

%-------------------
\subsection{Experimental constraints}
\label{sec:constraints}
The sum of neutrino masses, $\sum_im_i$, can be constrained by cosmological measurements. Combining the latest Planck CMB measurements with the baryon acoustic oscillation data, one obtains an upper bound $\sum_im_i\le 0.13-0.52$ eV at 95\% confidence level (CL)~\cite{Planck:2018vyg,DiValentino:2019dzu}. The range in the upper bound reflects the uncertainty in the assumed cosmological model. The upper bound on the neutrino masses sum translates to the bound on the lightest neutrino mass as
\begin{equation}
    m_{\mathrm{light}} \lesssim \left\{
    \begin{aligned}
        &34-171\text{ meV},\qquad\text{NO},\\
        &21-168\text{ meV},\qquad\text{IO}.
    \end{aligned} \right.
    \label{eq:cosmology}
\end{equation}
In the case of NO, we will refer to 34 (171) meV as the tight (loose) upper bound on the lightest neutrino mass. Similarly, for IO, the tight (loose) upper bound is 21 (168) meV.
%In our work, we will use 0.52 eV as a conservative upper bound.

The effective electron neutrino mass is defined as
\begin{equation}
    m_{\nu_e}^{\text{eff}} = \sqrt{\sum_i|U_{ei}|^2m_i^2}.
\end{equation}
It determines the end point of the beta decay spectrum. The strongest upper bound on $m_{\nu_e}^{\text{eff}}$ is provided by the KATRIN experiment. With the 259 days of tritium beta decay data, KATRIN obtains the upper bound $m_{\nu_e}^{\text{eff}}\le$ 0.45 eV at 90\% CL~\cite{Katrin:2024tvg}.

One of the hallmark of Majorana neutrino is the existent of the neutrinoless double beta decay ($0\nu\beta\beta$). Its rate is characterized by the $0\nu\beta\beta$ effective mass 
\begin{equation}
    m_{ee} = \left|\sum_i U_{ei}^2m_i\right| = \left|(m_\nu)_{ee}\right|.
\end{equation}
Currently, the strongest bound on $m_{ee}$ comes from the search for $0\nu\beta\beta$ in xenon by the KamLAND-Zen experiment with $m_{ee}\le 36-156$ meV at 90\% CL~\cite{KamLAND-Zen:2022tow}. The range in the upper bound is due to the uncertainties in computing nuclear matrix element. A slightly weaker bound, $m_{ee}\le 79-180$ meV at 90\% CL, is obtained by the GERDA experiment using germanium~\cite{GERDA:2020xhi}.

%-------------------
\subsection{Compatibility with the lightest neutrino mass being zero}
Current neutrino oscillation data allow for the lightest neutrino being massless. In this scenario, neutrino masses are completely determined from oscillation parameters. For example, in the case of NO one would get
\begin{equation}
    m_1 = 0,\quad
    m_2 = \sqrt{\Delta m^2_{\text{sol}}}\,,\quad
    m_3 = \sqrt{\Delta m^2_{\text{atm}}}\,.
\end{equation}
Moreover, for the effective mass $m_{\nu_e}^{\text{eff}}$, one gets 
\begin{equation}
    m_{\nu_e}^{\text{eff}} = \sqrt{s_{12}^2c_{13}^2\Delta m^2_{\text{sol}} + s_{13}^2\Delta m^2_{\text{atm}}}\,.
\end{equation}
Analogous expressions can be derived for the IO case.

It is interesting to determine which of the seven two-zero textures could accommodate the vanishing lightest neutrino mass. To this end, we make use of the basic invariants of the neutrino mass matrix in Eqs~\eqref{eq:trace}~--~\eqref{eq:det}. Recalling that, in our parametrization, the ratios $r_i>0$, one can readily verify that, for textures A1, A2, B3, and B4, $\det(m_\nu^\dagger m_\nu)$ is positive definite. Thus, they are not compatible with the lightest neutrino being massless. 

In the case of textures B1 and B2, one has
\begin{align}
    \det\left(m_\nu^{(\text{B1})\dagger} m_\nu^{(\text{B1})}\right) &=\det\left(m_\nu^{(\text{B2})\dagger} m_\nu^{(\text{B2})}\right)\nonumber\\&= \mu^6\left[(r_1r_2^2-r_3^2)^2 + 2r_1r_2^2r_3^2 (1+\cos\theta)\right].
\end{align}
The determinant is vanishing when $\theta=\pi$ and $r_3^2=r_1r_2^2$. Thus, in the limit where the lightest neutrino mass is vanishing there will be no $CP$ violation in the lepton mixing matrix. Such a scenario is compatible only with neutrino masses being NO. 
%In particular, in this case $m_2^2=\Delta m^2_{\text{sol}}$ and $m_3^2=\Delta m^2_{\text{atm}}$. 
Focusing on the two nonvanishing neutrino masses, Eqs.~\eqref{eq:cross} and~\eqref{eq:det} imply that
\begin{equation}
    r_1+r_2^2+r_1r_2^2 = 1.
\end{equation}
Applying the above relation, the resulting neutrino mass matrix reads
\begin{equation}
    \left.m_\nu^{\text{B1,B2}}\right|_{m_1=0} = \begin{pmatrix}
        -1 &r_2 &0\\ r_2&0 &r_2\sqrt{\frac{1-r_2^2}{1+r_2^2}}\\ 0&r_2\sqrt{\frac{1-r_2^2}{1+r_2^2}} &\frac{1-r_2^2}{1+r_2^2}
    \end{pmatrix}.
\end{equation}
Diagonalizing this mass matrix yields $s^2_{13}\sim1$ which is incompatible with the neutrino oscillation data. Thus, textures B1 and B2 cannot accommodate a vanishing neutrino mass.  

In the case of texture C, one has
\begin{align} 
    \det\left(m_\nu^{(\text{C})\dagger} m_\nu^{(\text{C})}\right) &= \mu^6 r_3^2\left[(2r_1r_2-r_3)^2 + 4r_1r_2r_3^2 (1-\cos\theta)\right].
\end{align}
For texture C, the lightest neutrino mass vanishes when $\theta = 0$ and $r_3 = 2r_1r_2$. Similar to textures B1 and B2, this implies no $CP$ violation in the lepton sector. Hence, the neutrino masses ordering must be NO. Applying Eqs.~\eqref{eq:cross} and~\eqref{eq:det} to the two nonvanishing neutrino masses, one gets
\begin{equation}
    r_2^2 + r_1^2 (1 + 4 r_2^2) = 2r_1r_2.
\end{equation}
The above equation does not admit a real solution. Hence, in texture C, the lightest neutrino cannot be massless. 

%%%%%%%%%%%%%%%%%%%%
\section{Numerical scan}
\label{sec:results}
In this section, we perform a scan on the two-zero texture parameter space. We take the following ranges for the free parameters:
\begin{equation}
    1 \text{ meV} \le \mu \le 1 \text{ eV},\quad
    10^{-3}\le r_i \le 10^3,\quad
    0\le\theta\le2\pi.
\end{equation}
The parameter $\mu$ determines the overall scale of neutrino mass matrix. Its range is chosen to cover the possible mass range of the heaviest neutrino ($m_3$ for NO and $m_2$ for IO). The range of $r_i$ is chosen so that $r_i$ is not too different from 1. If $r_i\ll1$ or $r_i\gg1$, at least one element of the neutrino mass matrix is negligible. Hence, the resulting mass matrix approximately contains three or more independent zeros and is incompatible with the oscillation data. We identify the region of parameter space which gives neutrino oscillation parameters within the 3$\sigma$ ranges in Table~\ref{table:data_nufit}. We then identify the viable range of $m_{\mathrm{light}}$, $m_{\nu_e}^{\mathrm{eff}}$, and $m_{ee}$ for each texture. Our results are collected in Table~\ref{tab:observables}. 

There are five free parameters for each texture compared to eight neutrino observables: the six oscillation parameters and the two effective masses.  Thus, one expects correlations among these observables. In fact, from our scan, we find strong correlations among $s_{23}^2$, $\delta$, $m_{\mathrm{light}}$, $m_{\nu_e}^{\mathrm{eff}}$, and $m_{ee}$ for each of the seven two-zero textures. 

\begin{table}[htbp]
\centering
\begin{tabular}{|c|c|c|c|c|c | c|}\hline
Texture & Ordering & $s^2_{23}$ &$\delta$ [$^\circ$]& $m_{\mathrm{light}}$ [meV] &$m^{\mathrm{eff}}_{\nu_e}$ [meV] &$m_{ee}$ [meV]\\\hline
A1 & NO & $\,0.43-0.58\,$ &$\,124-290\,$ &$4.3-7.5$ &$9.7-11.7$ & 0\\
A2& NO &$0.50-0.58$ &$236-290$ &$4.0-5.6$ &$9.7-10.4$ & 0\\\hline
\multirow{2}{*}{B1} & NO &$0.44-0.50$ &$268-270$ &$\ge57$ &$\ge57$ &$\ge57$ \\
& IO &$0.50-0.58$ &$\simeq270$ &$\ge48$ &$\ge68$ &$\ge68$\\\cline{2-7}
\multirow{2}{*}{B2} & NO &$0.50-0.58$ &$270-273$ &$\ge48$ &$\ge49$ &$\ge48$\\
& IO &$0.44-0.50$ &$\simeq270$ &$\ge60$ &$\ge78$ &$\ge78$\\\cline{2-7}
\multirow{2}{*}{B3} & NO &$0.44-0.50$ &$270-272$ &$\ge60$ &$\ge61$ &$\ge61$\\
& IO &$0.50-0.58$ &$269-270$ &$\ge51$ &$\ge71$ &$\ge70$\\\cline{2-7}
\multirow{2}{*}{B4} & NO &$0.50-0.58$ &$267-270$ &$\ge51$ &$\ge52$ &$\ge52$\\
& IO &$0.44-0.50$ &$\simeq270$ &$\ge64$ &$\ge80$ &$\ge80$\\\hline
\multirow{2}{*}{C} & NO &$0.50$ &$124-289$ &$\ge156$ &$\ge156$ &$\ge149$\\
& IO &$0.44-0.58$ &$238-291$ &$\ge43$ &$\ge65$ &$\ge 43$\\\hline
\end{tabular}
\caption{The range of $m_{\mathrm{light}}$, $m_{\nu_e}^{\mathrm{eff}}$ and $m_{ee}$ in each scenario.}
\label{tab:observables}
\end{table}

%-------------------------
\subsubsection{Textures A1 and A2}

%\begin{figure}[ht!]
%\centering
%\includegraphics[width=0.45\textwidth]{Ar1vss23.pdf}
%\includegraphics[width=0.45\textwidth]{Athetavsdelta.pdf}
%\caption{The relation between the parameter $r_1$ and $s^2_{23}$ (left) and $\theta$ and Dirac phase $\delta$ (right) in texture A1 and A2. }
%\label{fig:Aparam}
%\end{figure}
\begin{figure}[ht!]
\centering
\includegraphics[width=0.45\textwidth]{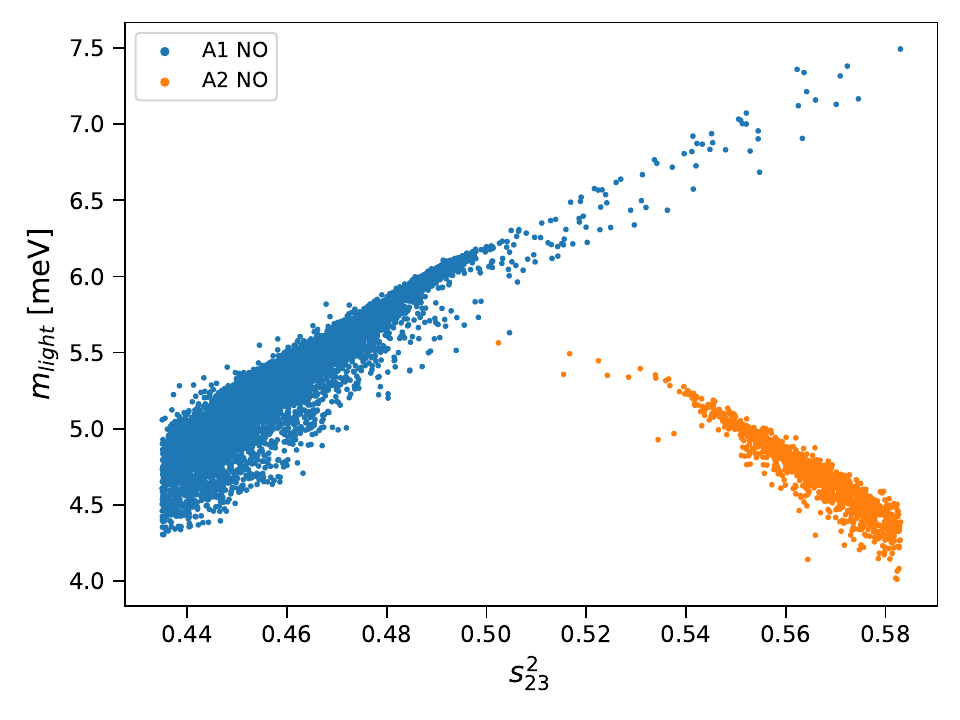}
\includegraphics[width=0.45\textwidth]{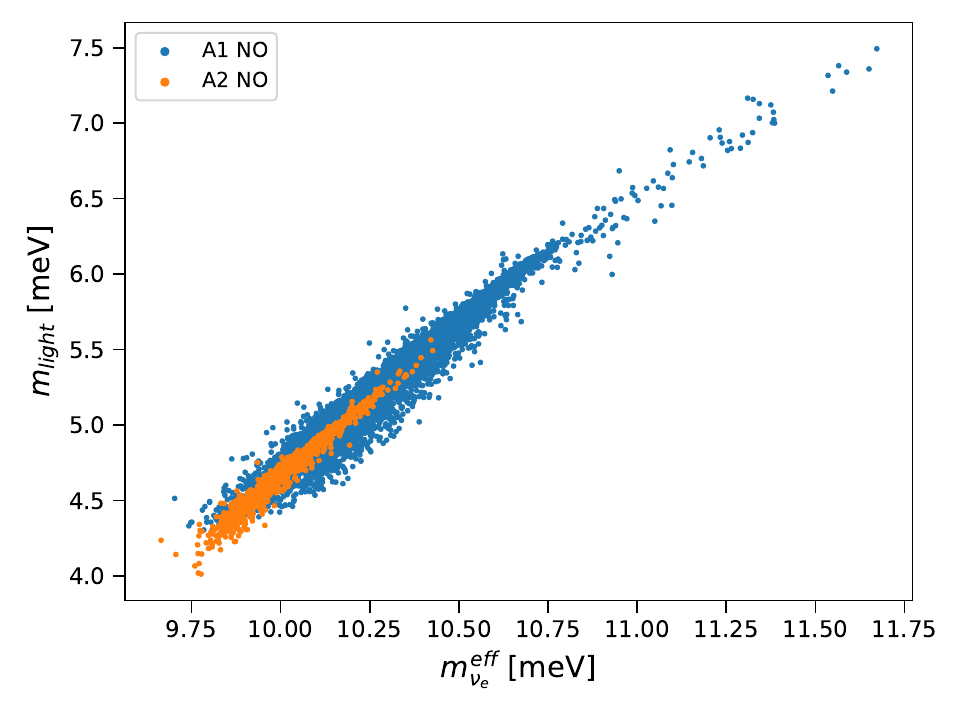}
\caption{The relation between $s^2_{23}$ and $m_{\mathrm{light}}$ (left) and $m_{\nu_e}^{\mathrm{eff}}$ and $m_{\mathrm{light}}$ (right) in textures A1 and A2. }
\label{fig:Amasses}
\end{figure}

\begin{figure}[ht!]
\centering
\includegraphics[width=0.45\textwidth]{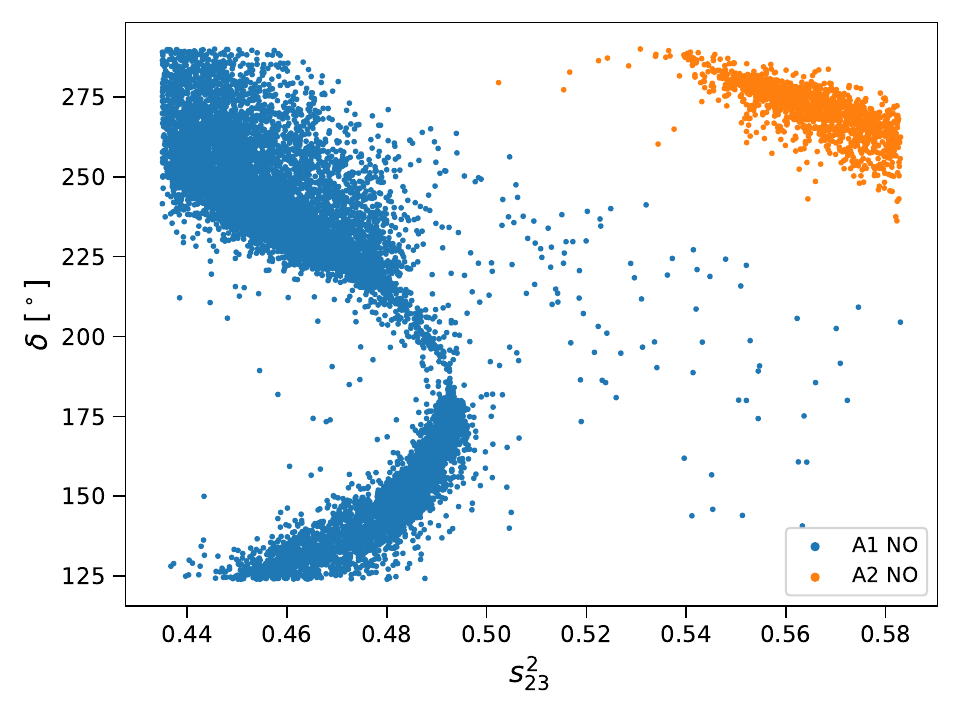}
\includegraphics[width=0.45\textwidth]{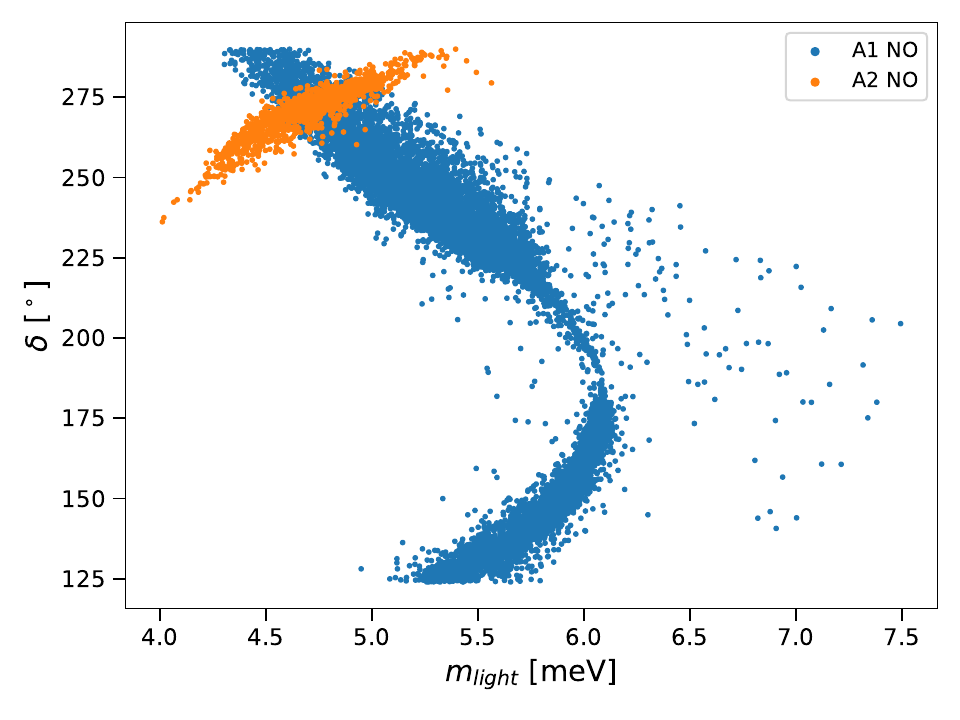}
\caption{The relation between $s^2_{23}$ and $\delta$ (left) and $m_{\mathrm{light}}$ and $\delta$ (right) in textures A1 and A2. }
\label{fig:Adelta}
\end{figure}

From our analysis, textures A1 and A2 can accommodate only NO neutrino masses. We find, in both cases, strong correlations between $s_{23}^2$, $m_{\mathrm{light}}$, and $m_{\nu_e}^{\mathrm{eff}}$; see Fig.~\ref{fig:Amasses}. For texture A1, $\theta_{23}$ can be in either the first and the second octant, while for A2, $\theta_{23}$ lies exclusively in the second octant. We also find weaker correlations between $s_{23}^2$, $m_{\mathrm{light}}$, and $\delta$; see Fig.~\ref{fig:Adelta}. It is interesting to note that large $CP$ violation ($\delta\simeq3\pi/2$) is associated with a small $m_{\mathrm{light}}$ value.

%the parameter $r_1$ is strongly correlated with $s^2_{23}$, and the parameter $\theta$ is correlated with the Dirac phase $\delta$, see Fig.~\ref{fig:Aparam}. It is interesting to note that for texture A2, $s^2_{23}>0.5$ and $\theta_{23}$ lies in the second octant.

Textures A1 and A2 are quite predictive in terms of neutrino masses. The lightest neutrino mass lie in a small window with 4.3 meV $\lesssim m_{\mathrm{light}}^{(\mathrm{A1})}\lesssim 7.5$ meV and 4.0 meV $\lesssim m_{\mathrm{light}}^{(\mathrm{A2})}\lesssim 5.5$ meV. This range is roughly an order of magnitude below the upper bound from cosmology, $m_{\mathrm{light}}\le34-171$ meV. The lightest neutrino mass is also correlated with $s^2_{23}$; see Fig.~\ref{fig:Amasses}. The effective mass $m_{\nu_e}^{\mathrm{eff}}$, similar to the lightest neutrino mass, lies in a small window with 9.7 meV $\lesssim m_{\nu_e}^{\mathrm{eff}}\lesssim 11.7$ meV (9.7 meV $\lesssim m_{\nu_e}^{\mathrm{eff}}\lesssim 10.4$ meV) for texture A1 (A2). Again, as in the case of the lightest neutrino mass, this mass window is an order of magnitude below the KATRIN constraint, {\color{blue} $m_{\nu_e}^{\text{eff}}\le450$} meV. The effective mass $m_{ee}$, on the other hand, is vanishing for both A1 and A2.

\subsubsection{Textures B1--B4}

\begin{figure}[ht!]
\centering
\includegraphics[width=0.45\textwidth]{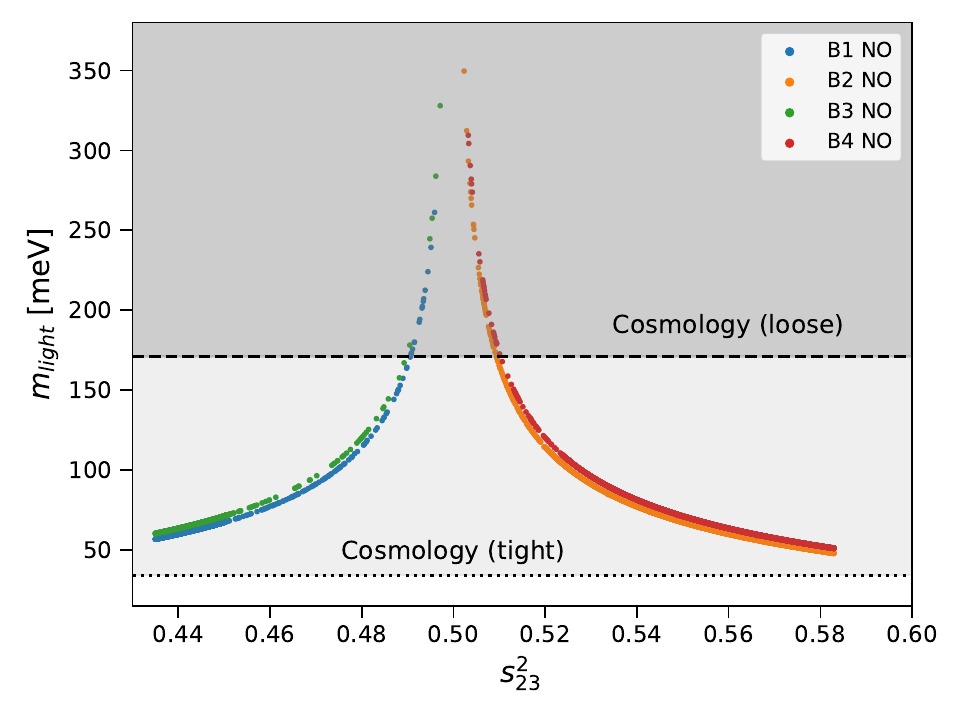}
\includegraphics[width=0.45\textwidth]{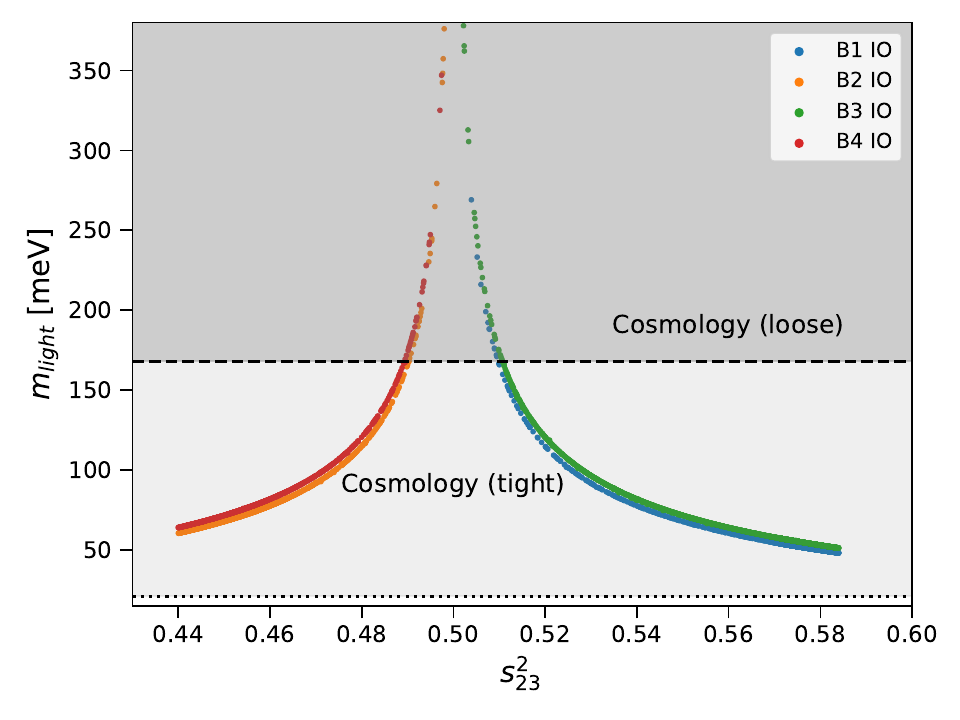}
\caption{The relation between $s^2_{23}$ and the lightest neutrino mass $m_{\mathrm{light}}$ for NO (left) and IO (right). The dotted (dashed) line shows the loose (tight) upper bound on $m_{\mathrm{light}}$ from cosmological measurements.}
\label{fig:Bs23vsmlight}
\end{figure}

\begin{figure}[ht!]
\centering
\includegraphics[width=0.45\textwidth]{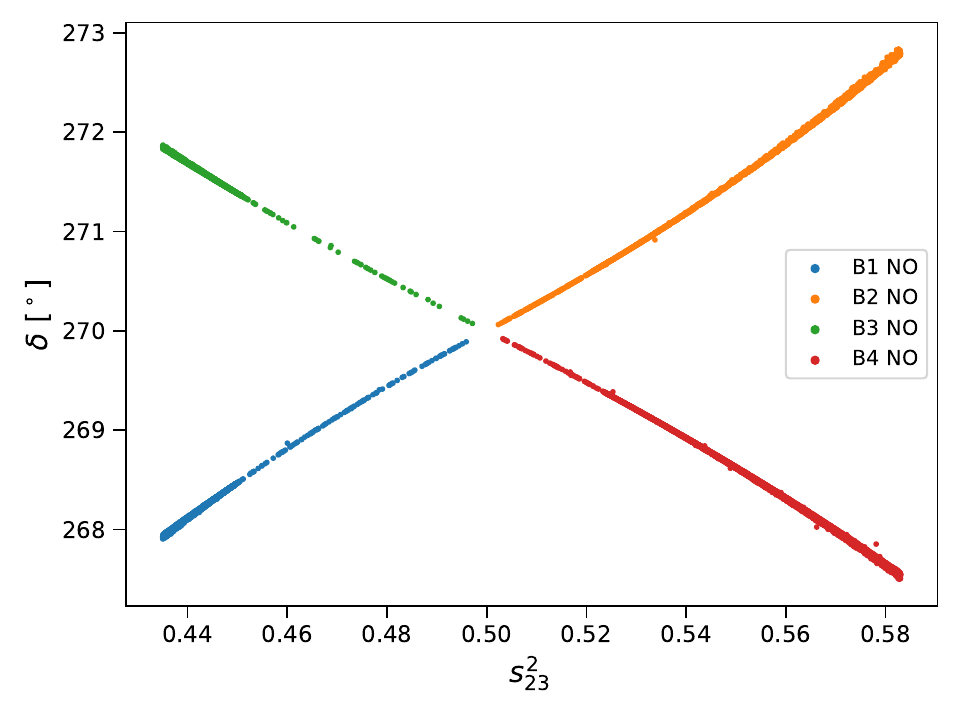}
\includegraphics[width=0.45\textwidth]{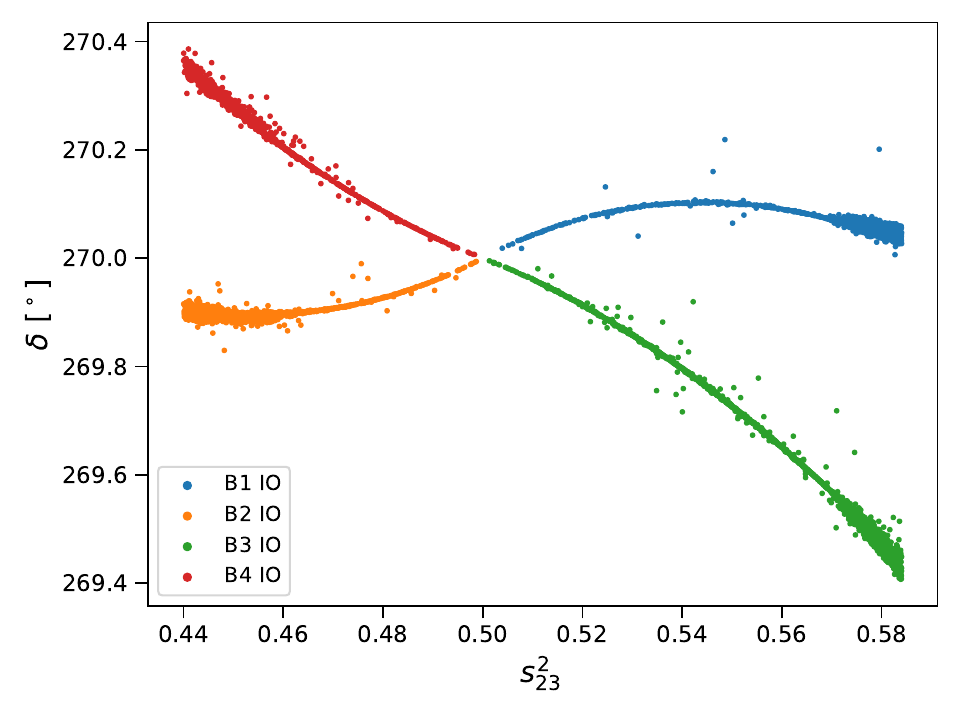}
\caption{The relation between $s^2_{23}$ and $\delta$ for NO (left) and IO (right). }
\label{fig:Bs23vsdelta}
\end{figure}

From our scan, textures B1--B4 are compatible with both NO and IO neutrino masses. As in the cases of textures A1 and A2, here we find a strong correlation between $s_{23}^2$ and $m_{\mathrm{light}}$; see Fig.~\ref{fig:Bs23vsmlight}. However, unlike in A1 and A2, $s_{23}$ is also strongly correlated with $\delta$; see Fig.~\ref{fig:Bs23vsdelta}. It is interesting to note that, for both NO and IO scenarios of textures B1--B4, the values of $\delta$ are clustered near the maximal $CP$ violation value of $\delta = 3\pi/2$. 

In the case of B1 and B3, $\theta_{23}$ lies entirely in the first (second) octant for the NO (IO) scenario. This corresponds to $s_{23}^2$ being in the vicinity of the true minimum of the global fit to neutrino data~\cite{Esteban:2024eli}. In the cases of B2 and B4, on the other hand, $s_{23}^2$ lies in the vicinity of the local minimum of the fit. It is interesting to note that the lightest neutrino mass increases as $s^2_{23}$ approaches 0.5 from above and below.

Our scan also reveals that, for textures B1--B4, $m_{\mathrm{light}}\simeq m_{\nu_e}^{\mathrm{eff}}\simeq m_{ee}$ in the NO scenario and $m_{\mathrm{light}}< m_{\nu_e}^{\mathrm{eff}}\simeq m_{ee}$ in the IO scenario. In particular, in the NO scenario we find $m_{\mathrm{light}}\gtrsim 57$ meV for textures B1 and B3 and $m_{\mathrm{light}}\gtrsim 48$ meV for textures B2 and B4. In the IO scenario, we find $m_{\mathrm{light}}\gtrsim 48$ meV for textures B1 and B3 and $m_{\mathrm{light}}\gtrsim 61$ meV for textures B2 and B4. Our lower bounds on $m_{\mathrm{light}}$ are incompatible with the tight upper bound from cosmology; see Sec.~\ref{sec:constraints}. However, they are still allowed by the loose upper bound. For the $0\nu\beta\beta$ effective mass $m_{ee}$ in the IO scenario, we find $m_{ee}\gtrsim68$ meV for textures B1 and B3 and $m_{ee}\gtrsim78$ meV for textures B2 and B4. The ranges of $m_{ee}$ for textures B1--B4 can be probed by the $0\nu\beta\beta$ experiments. In particular the KamLAND-Zen experiment can probe $m_{ee}$ down to the range $36-156$ meV. The effective mass $m_{\nu_e}^{\mathrm{eff}}$, on the other hand, lies well below the sensitivity of the current KATRIN experiment. 

\subsubsection{Texture C}

\begin{figure}[ht!]
\centering
\includegraphics[width=0.45\textwidth]{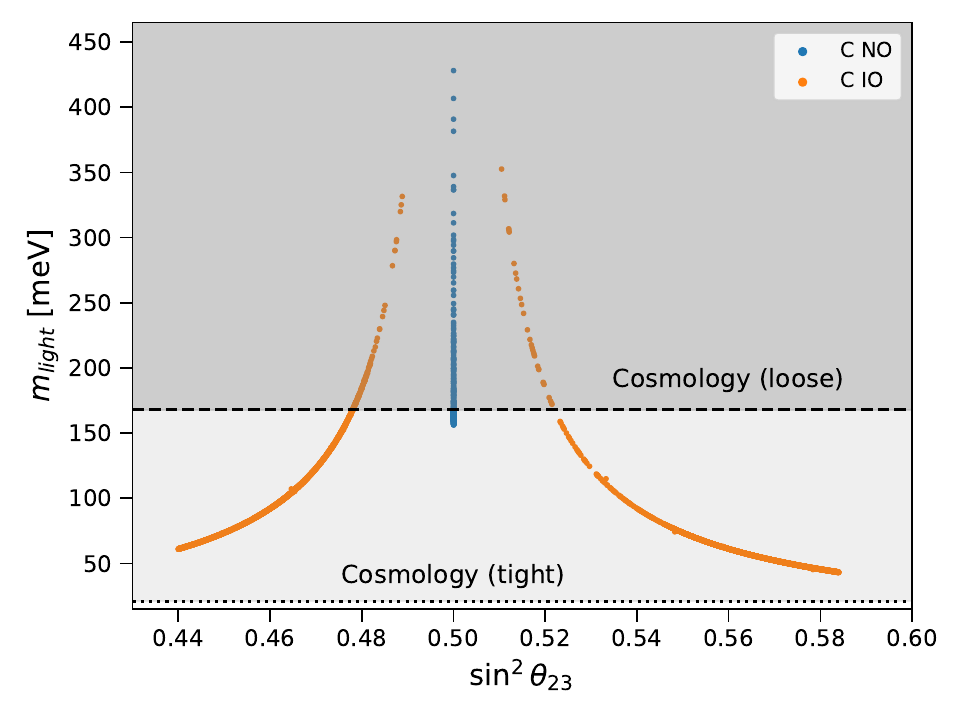}
\includegraphics[width=0.45\textwidth]{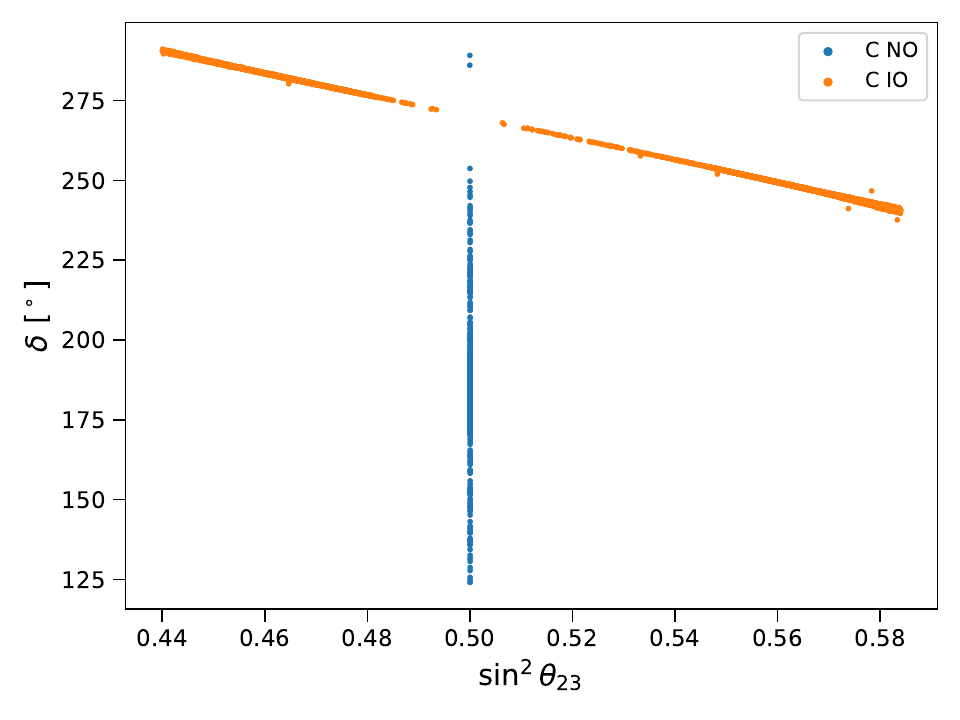}
\caption{The relation between $s^2_{23}$ and $m_{\mathrm{light}}$ (left) and $s_{23}^2$ and $\delta$ (right) in texture C. The dotted (dashed) line shows the loose (tight) upper bound on $m_{\mathrm{light}}$ from cosmological measurements.}
\label{fig:Cobserables}
\end{figure}

Texture C can accommodate both NO and IO neutrino masses. For NO, we find that $s^2_{23}\simeq0.5$ and neutrino masses are quasidegenerate, while for IO $\theta_{23}$ can be in either octant; see Fig.~\ref{fig:Cobserables}. The Dirac phase $\delta$ can take any value between 124$^\circ$ and $289^\circ$ in the NO scenario. In the IO case, $\delta$ is clustered around 270$^\circ$. It is interesting to note that, in the IO scenario, the lightest neutrino mass increases as $s^2_{23}$ approaches 0.5 from above and below; see Fig.~\ref{fig:Cobserables}.

Texture C is in severe tension with a cosmological constraint on neutrino mass. For the NO scenario, we find $m_{\mathrm{light}}\gtrsim156$ meV, with only a small portion of parameter space consistent with the loose upper bound on the lightest neutrino mass, $m_{\mathrm{light}}\lesssim171$ meV. However, this parameter space is completely ruled out by the tight upper bound $m_{\mathrm{light}}\lesssim34$ meV. On the other hand, for the IO scenario, we find $m_{\mathrm{light}}\gtrsim43$ meV, leaving a large part of the parameter space consistent with the loose upper bound $m_{\mathrm{light}}\lesssim168$ meV. Nevertheless, it is also ruled out by the tight upper bound $m_{\mathrm{light}}\lesssim21$ meV.

Similar to textures B1--B4, our scan shows that $m_{\nu_e}^{\mathrm{eff}}$ and $m_{ee}$ are strongly correlated with $m_{\mathrm{light}}$. In particular, we find $m_{ee} < m_{\nu_e}^{\mathrm{eff}}\simeq m_{\mathrm{light}}$ for the NO scenario with $m_{ee}\gtrsim149$ meV.  For the IO scenario, we find $m_{ee}\simeq m_{\mathrm{light}}<m_{\nu_e}^{\mathrm{eff}}$ with $m_{\nu_e}^{\mathrm{eff}}\gtrsim65$ meV. The difference between $m_{\mathrm{light}}$ and $m_{ee}$ ($m_{\nu_e}^{\mathrm{eff}}$) in the NO (IO) case decreases as $m_{\mathrm{light}}$ increases. Note that, in both NO and IO scenarios, the $0\nu\beta\beta$ effective mass $m_{ee}$ is in tension with the KamLAND-Zen constraint $m_{ee}<36-156$ meV, while the effective mass $m_{\nu_e}^{\mathrm{eff}}$ remains well below the reach of the KATRIN experiment.

%%%%%%%%%%%%%%%%%%%%
\section{Conclusion and discussion}
\label{sec:conclusion}
We have revisited the two-zero texture Majorana neutrino mass matrix. In the literature, seven two-zero textures [see Eq.~\eqref{eq:textureA}~--~\eqref{eq:textureC}] have been identified as being consistent with the measured neutrino oscillation parameters. In this work, we subject the seven two-zero textures to the current neutrino data. We show that none of the seven textures can accommodate the lightest neutrino mass being zero. Our analysis also shows that textures B1 -- B4 and C are in tension with constraints on neutrino mass from cosmology and $0\nu\beta\beta$ experiments. A similar conclusion has also been reached in Ref.~\cite{Denton:2023hkx}.

For textures A1 and A2, neutrino masses are NO. The sum of neutrino masses lies in a small mass window with 64 meV $\lesssim \sum_i m_i^{(\mathrm{A1})}\lesssim 70$ meV and 64 meV $\lesssim \sum_im_{i}^{(\mathrm{A2})}\lesssim 66$ meV. These mass windows are an order of magnitude below the current experimental sensitivity. Nevertheless, they are within reach of the Simon Observatory with a projected sensitivity on the total neutrino mass of 40 meV~\cite{SimonsObservatory:2019qwx}. The effective mass $m_{\nu_e}^{\mathrm{eff}}$, on the other hand, lies in the range $9.6-10.4$ meV. This is an  order of magnitude below the projection $m_{\nu_e}^{\mathrm{eff}}\lesssim200$ meV at 90\% C.L. of both the HOLMES and the KATRIN experiments~\cite{Alpert:2014lfa,KATRIN:2019yun}.

%also need to mention s_{23} and A2

Textures B1--B4 and C can accommodate both NO and IO. All of them are in tension with the constraints on the neutrino mass sum (or, equivalently, the lightest neutrino mass) and the effective mass for $0\nu\beta\beta$ decay. In all cases except texture C with NO, there is a large part of parameter space compatible with the loose constraint $\sum_im_i\lesssim 520$ meV. However, this part of parameter space is disallowed by the tight constraint $\sum_im_i\lesssim 130$ meV. The Simon Observatory measurement could decisively rule out textures B1--B4 and C. Moreover, the value of $m_{ee}$ in these texture lies in the range that is probed by the KamLAND-Zen experiment. The planned phase II of the LEGEND experiment, with the projected upper bound $m_{ee}\le13-29$ meV~\cite{LEGEND:2017cdu}, could also decisively rule out textures B1--B4 and C. The effective mass $m_{\nu_e}^{\mathrm{eff}}$, on the other hand, lies below the projected sensitivity of both the HOLMES and the KATRIN experiments.

The mixing angle $s_{23}^2$ and the neutrino mass ordering can also play a crucial role in probing the two-zero textures neutrino mass matrix. If $\theta_{23}$ is decisively determined to be in the first octant for the NO scenario, as currently preferred by the global neutrino data fit~\cite{Esteban:2024eli}, textures A2, B2, B4 and C would be ruled out. Similarly, if $\theta_{23}$ is determined to be in the second octant for the IO case, textures B2 and B4 would be excluded. The neutrino mass ordering could be determined by the upcoming JUNO experiment~\cite{JUNO:2015zny,JUNO:2021vlw}. The octant of $\theta_{23}$, in principle, could be determined by the next generation long-baseline neutrino experiments such as DUNE~\cite{DUNE:2020jqi} and Hyper-Kamiokande~\cite{Hyper-Kamiokande:2018ofw}.

\begin{acknowledgments}
N. S. and P. U. acknowledge support from the National Science, Research and Innovation Fund (NSRF) for the fiscal year 2025 via Srinakharinwirot University  under Grant No. 046/2568. W.T. acknowledges support from the National Research Council of Thailand (NRCT): NRCT5-RGJ63017-153. P.U. also thanks the High-Energy Physics Research Unit, Chulalongkorn University for the hospitality while part of this work is being completed. The authors acknowledge the National Science and Technology Development Agency, National e-Science Infrastructure Consortium, Chulalongkorn University and the Chulalongkorn Academic Advancement into Its 2nd Century Project, NSRF via the Program Management Unit for Human Resources and Institutional Development, Research and Innovation [Grant No. B05F650021 and No. B37G660013] (Thailand) for providing computing infrastructure that has contributed to the research results reported within this paper.
\end{acknowledgments}

%%%%%%%%%%%%%%%%%%%%

%%%%%%%%%%%%%%%%%%%%
\bibliographystyle{apsrev4-1}
\bibliography{reference}
\end{document}